\documentclass[11pt]{iopart}
\usepackage{iopams}
\expandafter\let\csname equation*\endcsname\relax
\expandafter\let\csname endequation*\endcsname\relax
\usepackage{amsmath}
\usepackage{amsfonts}
\usepackage{amsthm}
\usepackage{dsfont}
\usepackage{graphicx}

\def\Im{{\rm Im}}

\newcommand{\rf}[1]{(\ref{#1})}

\begin{document}
\title[]{Stationary states of a $PT$-symmetric two-mode Bose-Einstein condensate}
\author{Eva-Maria Graefe}
\address{Department of Mathematics, Imperial College London, London, SW7 2AZ, United Kingdom}
\date{\today}

\begin{abstract} 
The understanding of nonlinear $PT$-symmetric quantum systems, arising for example in the theory of Bose-Einstein condensates in $PT$-symmetric potentials, is widely based on numerical investigations, and little is known about generic features induced by the interplay of $PT$-symmetry and nonlinearity. To gain deeper insights it is important to have analytically solvable toy-models at hand. In the present paper the stationary states of a simple toy-model of a $PT$-symmetric system are investigated. The model can be interpreted as a simple description of a Bose-Einstein condensate in a $PT$-symmetric double well trap in a two-mode approximation. The eigenvalues and eigenstates of the system can be explicitly calculated in a straight forward manner; the resulting structures resemble those that have recently been found numerically for a more realistic $PT$-symmetric double delta potential. In addition, a continuation of the system is introduced that allows an interpretation in terms of a simple linear matrix model.
\end{abstract}

\pacs{03.65.Ge, 03.75.Hh}

%\maketitle

\section{Introduction}
Recent years have seen an increasing interest in non-Hermitian, and in particular $PT$-symmetric quantum theories \cite{Mois11book,Bend99a}, which concern quantum systems with complex Hamiltonians that nevertheless can have purely real eigenvalues. $PT$-symmetric quantum theories with a purely real spectrum (the case of so-called unbroken $PT$ symmetry), have been suggested as generalisations of Hermitian quantum theory on a fundamental level \cite{Bend02b}. Considerable work has been devoted to the introduction of a theoretical framework that leads to unitary time evolutions and physically meaningful concepts of observables in these systems (see, e.g., \cite{Bend02b,PTtheo} and references therein). The focus of research in $PT$-symmetric quantum theories has shifted recently, with the proposal and development of experimental applications, where $PT$-symmetric quantum theories arise as effective non-Hermitian descriptions of certain open quantum systems, that is, systems in the presence of absorption and gain (see, e.g. \cite{nherm_optics} and references therein). There has been considerable progress in this area, and the first $PT$-symmetric systems have been experimentally realised in optical waveguide structures \cite{Guo09,Ruet10}, microwave cavities \cite{Bittner} and electronic circuits \cite{Kottos11}. While there have been a number of proposals for the realisation of $PT$-symmetric theories in \textit{bona fide} quantum systems, such as Bose-Einstein condensates (BECs) in optical lattices \cite{08nhbh_s,nhbh,Kevr,08PT,Muss08,PT_BEC,Wunn12}, so far these models have not been experimentally realised. 

The consideration of BECs adds an additional interesting feature to the standard models of $PT$-symmetric quantum systems, as the effective description of BECs via a mean-field theory is typically nonlinear. Other fields in which nonlinear $PT$-symmetric problems have been investigated are classical wave systems \cite{Cavaglia}, and complex nonlinear Schr\"odinger dynamics arising in the context of nonlinear optics \cite{Muss08,Ramezani}. The interest in the field is still relatively new, and we are only beginning to understand the peculiarities of nonlinear versions of $PT$-symmetric systems. One main hinderance is the lack of analytically tractable systems and toy models in which generic properties can be analysed and investigated. 

In the present paper we present an analytically solvable nonlinear $PT$ symmetric eigenvalue problem that models stationary states of a Bose-Einstein condensate in a $PT$-symmetric two-mode potential. This is the stationary version of a model introduced in \cite{08nhbh_s,nhbh} as the mean-field limit of a $PT$-symmetric two-mode many-boson system \cite{08PT,Hill06}. In \cite{08nhbh_s,nhbh} the dynamical properties have been investigated in detail. Here we extend these results to give \textit{analytic} expressions for the corresponding nonlinear eigenvalues, and analyse the breakdown of $PT$-symmetry in this nonlinear model. Eigenvalue structures similar to the ones obtained here have recently been found numerically in a model of a Bose-Einstein condensate in a double-delta potential \cite{Wunn12}. Similar models have also been investigated in the context of two coupled nonlinear wave guides \cite{Ramezani}. 

A typical feature of nonlinearity is that it can lead to additional stationary states. In the model investigated here, there are up to four eigenstates for arbitrary parameter values \cite{nhbh}, while there are only two eigenstates in the linear limit. Allowing for suitable complex extensions of the system, four eigenstates can be found for almost all parameter values, with the exception of isolated exceptional points at which two stationary states coalesce \cite{Mois11book,EP}. These four eigenvalues can be interpreted as those of a linear $4\times 4$ matrix, which we give explicitly. We furthermore show that there is also a direct relation between the stationary states of the nonlinear $2\times 2$ and the linear $4\times4$ system. 

The paper is organised as follows: In Section \ref{sec_model} we introduce the model system and contrast it to similar models considered in the literature. In Section \ref{sec-stat} we investigate the corresponding time-independent system, provide analytic expressions for the eigenstates and eigenvalues, and investigate the stability of the stationary solutions. In Section \ref{sec_iso} we introduce an analytic continuation of the model that has four eigenstates for all parameter values and show that it can be interpreted in terms of a linear matrix model. We conclude in Section \ref{sec_sum} with a summary and outlook.

\section{The model system}
\label{sec_model}
For low temperatures, the dynamics of a BEC in a real-valued potential $V(x)$ can be described
in a mean-field approach by the following nonlinear Schr\"odinger equation (NLSE) of Gross-Pitaevskii type \cite{Pita03}
\begin{equation}
  \left( - \frac{\hbar^2}{2m} \frac{\partial^2}{\partial x^2} + V(x) + g |\psi(x,t)|^2 \right)
  \psi(x,t) =\rmi \hbar \frac{\partial \psi(x,t)}{\partial t}.
  \label{eqn-NLSE-timedep}
\end{equation}
The coefficient of the nonlinear term is given by $g=4\pi\hbar^2a/m$, where $a$ is the s-wave scattering length. 

In the case of a double well potential for low energies it is often 
appropriate to approximate the wave function by a two-mode
model by writing 
$$\psi(x,t) = \psi_1(t) \xi_1(x) + \psi_2(t) \xi_2(x),$$ 
where $\xi_{1,2}$ are localised in the left or the right side 
of the double well, respectively (see, e.g., \cite{Milb97}). 
Neglecting the nonlinear coupling of the two modes, one obtains the two-state model
\begin{equation}
 \rmi \hbar \frac{\rmd}{\rmd t} \left(\begin{array}{c} \psi_1 \\ \psi_2 \\ \end{array}\right)
  = \left(\begin{array}{c c} g|\psi_1|^2 & v \\
v & g|\psi_2|^2 \\
\end{array}\right)\left(\begin{array}{c} \psi_1 \\ \psi_2 \\ \end{array}\right)=
  H(|\psi_1|^2,|\psi_2|^2) \left(\begin{array}{c} \psi_1 \\ \psi_2 \\ \end{array}\right),
  \label{eqn-2state-model}
\end{equation}
with a Hamiltonian matrix $H(|\psi_1|^2,|\psi_2|^2)$ that depends on the population
of the two modes. The parameter $v$ is related to the height of the barrier between the two wells, which are assumed to be symmetric here. Note that the same Hamiltonian also appears in a variety of other physical systems, including nonlinear optics and molecular dynamics \cite{Kenk86,Eilb85}.

In the context of BECs, the Gross-Pitaevskii equation is an effective single particle approximation of a many-particle quantum system of cold atoms. Although it has been introduced formally only for real potentials, recently there has been considerable interest in complex generalisations of Gross-Pitaevskii equations \cite{GP_compl,06nlnh,07nlres,09ddshell,Wunn12,Kevr}. Considering in particular potentials whose real part is symmetric and whose imaginary part anti-symmetric, $V(-x)=V^*(x)$, one obtains $PT$-symmetric versions of the Gross-Pitaevskii equation\footnote{Strictly speaking, the $PT$-symmetry of the system is only given, if the wave-function respects the symmetry and thus renders the nonlinear part in the equation $PT$-symmetric.}.  In this spirit, a BEC in a $PT$-symmetric two-mode potential is described by a nonlinear Hamiltonian of the form
\begin{equation}
H = \left(\begin{array}{c c} g|\psi_1|^2 -\rmi\gamma & v \\
v & g|\psi_2|^2 +\rmi\gamma \\
\end{array}\right).
\label{eqn-nlin-nherm-ham-intro}
\end{equation}
This system is $PT$ symmetric with respect to a parity operator $P$ that interchanges the two basis states, $P: \psi_1\leftrightarrow \psi_2$, and the standard time reversal operator $T$ of complex conjugation, $T: \rmi \to -\rmi$. The physical interpretation of the complex energies is a loss of particles in the side with negative imaginary part, and an inflow of particles in the side with positive imaginary part. The model can be transformed to a purely lossy non-Hermitian model via a constant imaginary energy shift \cite{08nhbh_s,nhbh,Guo09}. The Hamiltonian \rf{eqn-nlin-nherm-ham-intro} also appears in the description of $PT$ symmetric wave guide structures with Kerr nonlinearity \cite{Ramezani}. 

The Gross-Pitaevskii equation, however, is an effective description of a many-particle system, valid if the system is condensed; and it is \textit{a priori} not clear whether this so-called mean-field approximation commutes with the complexification of the system. The mean-field approximation of non-Hermitian many-particle systems is closely related to the classical limit of non-Hermitian single-particle systems, which itself is a topic of recent research \cite{09nhclass,GraeSchu11,Bend10}. In the investigation of example systems it has been shown that these two limits do not commute in general. In \cite{08nhbh_s,nhbh} the mean-field approximation of a two-mode many-particle system with complex energies has been introduced, and it has been found that due to the non-conservation of the total probability in the presence of complex energies\footnote{Even if the system is in the unbroken $PT$-symmetric phase the dynamics of an arbitrary initial state that is not an eigenstate does not conserve the norm.}, the nonlinear term in the resulting Gross-Pitaevskii equation is modified. Instead of the system \rf{eqn-nlin-nherm-ham-intro} one finds the dynamical equation
\begin{equation}\label{nlnhGP}
\rmi\frac{\rmd}{\rmd\, t}\begin{pmatrix} {\psi}_1 \\{\psi}_2 \end{pmatrix}
=\begin{pmatrix}    g\,\frac{|\psi_1|^2}{|\psi_1|^2+|\psi_2|^2}  - \rmi \gamma & v \\
  v &   g\, \frac{|\psi_2|^2}{|\psi_1|^2+|\psi_2|^2} + \rmi \gamma \end{pmatrix}\begin{pmatrix}\psi_1\\ \psi_2\end{pmatrix}.
  \end{equation}
Since the norm $n=|\psi_1|^2+|\psi_2|^2$ is time dependent for non-vanishing $\gamma$, the dynamics resulting from \rf{nlnhGP} is fundamentally different from the dynamics generated by complex Gross-Pitaevskii equations of the form \rf{eqn-nlin-nherm-ham-intro}. Details of the dynamical features of \rf{nlnhGP} can be found in \cite{08nhbh_s,nhbh}, whereas the dynamics of \rf{eqn-nlin-nherm-ham-intro} is discussed in \cite{Ramezani}. Obviously \rf{nlnhGP} and \rf{eqn-nlin-nherm-ham-intro} coincide in both the linear and the Hermitian limit. Further, the time-independent versions of these two complex nonlinear Schr\"odinger equations only differ by a constant scaling of the nonlinearity parameter. Thus, when investigating stationary states the systems \rf{nlnhGP} and \rf{eqn-nlin-nherm-ham-intro} are equivalent.

\section{Stationary states and nonlinear eigenvalues}
\label{sec-stat}
In this section we analyse the stationary solutions of the form $\vec{\psi}(t)=\rme^{-i\mu t}\vec{\phi}$
for the nonlinear complex Schr\"odinger equation \rf{nlnhGP}, and the corresponding 
generalised nonlinear eigenvalues $\mu$. Assuming $\vec{\phi}$ to be normalised to unity this leads to the time independent complex nonlinear Schr\"odinger equation 
\begin{equation}
\label{complextigpe_a}
\begin{pmatrix} - \rmi \gamma + g |\phi_1|^2  & v \\
  v &  \rmi \gamma +g |\phi_2|^2 \end{pmatrix} \left(\begin{array}{c} \phi_1 \\ \phi_2 \\ \end{array}\right)=\mu \left(\begin{array}{c} \phi_1 \\ \phi_2 \\ \end{array}\right),
 \end{equation}
with chemical potential $\mu$, which can become complex in the present case. 
To simplify the calculation we introduce the constant energy shift $\frac{g}{2}(|\phi_1|^2+|\phi_2|^2)$, leading to the eigenvalue equation
\begin{equation}
\label{complextigpe}
\begin{pmatrix} - \rmi \gamma + c( |\phi_1|^2-|\phi_2|^2)  & v \\
  v &  \rmi \gamma -c (|\phi_1|^2 -|\phi_2|^2) \end{pmatrix} \left(\begin{array}{c} \phi_1 \\ \phi_2 \\ \end{array}\right)=\mu \left(\begin{array}{c} \phi_1 \\ \phi_2 \\ \end{array}\right),
 \end{equation}
with $c=g/2$. 

\begin{figure}[tb]
\begin{center}
\includegraphics[width=5cm]{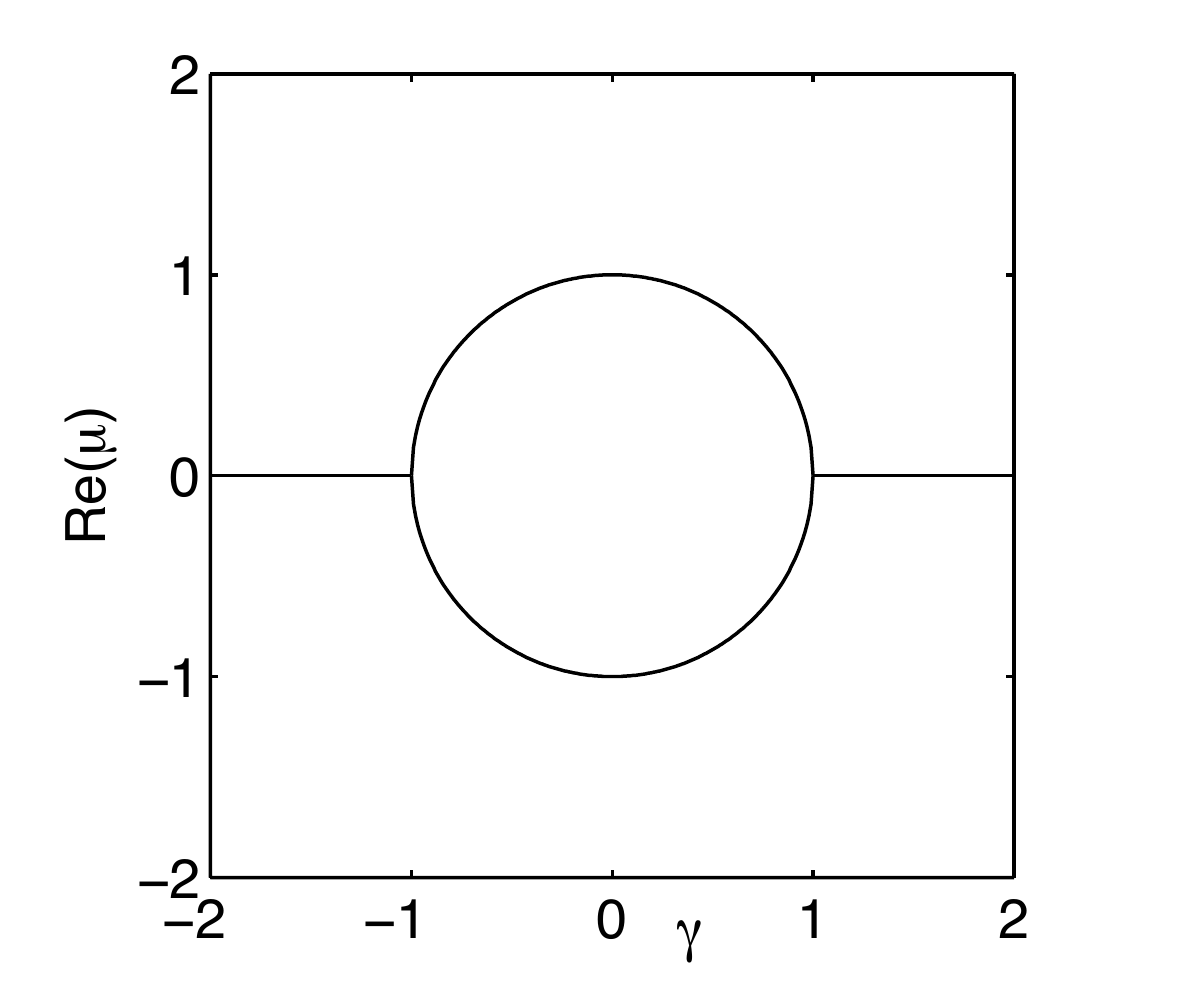}
\includegraphics[width=5cm]{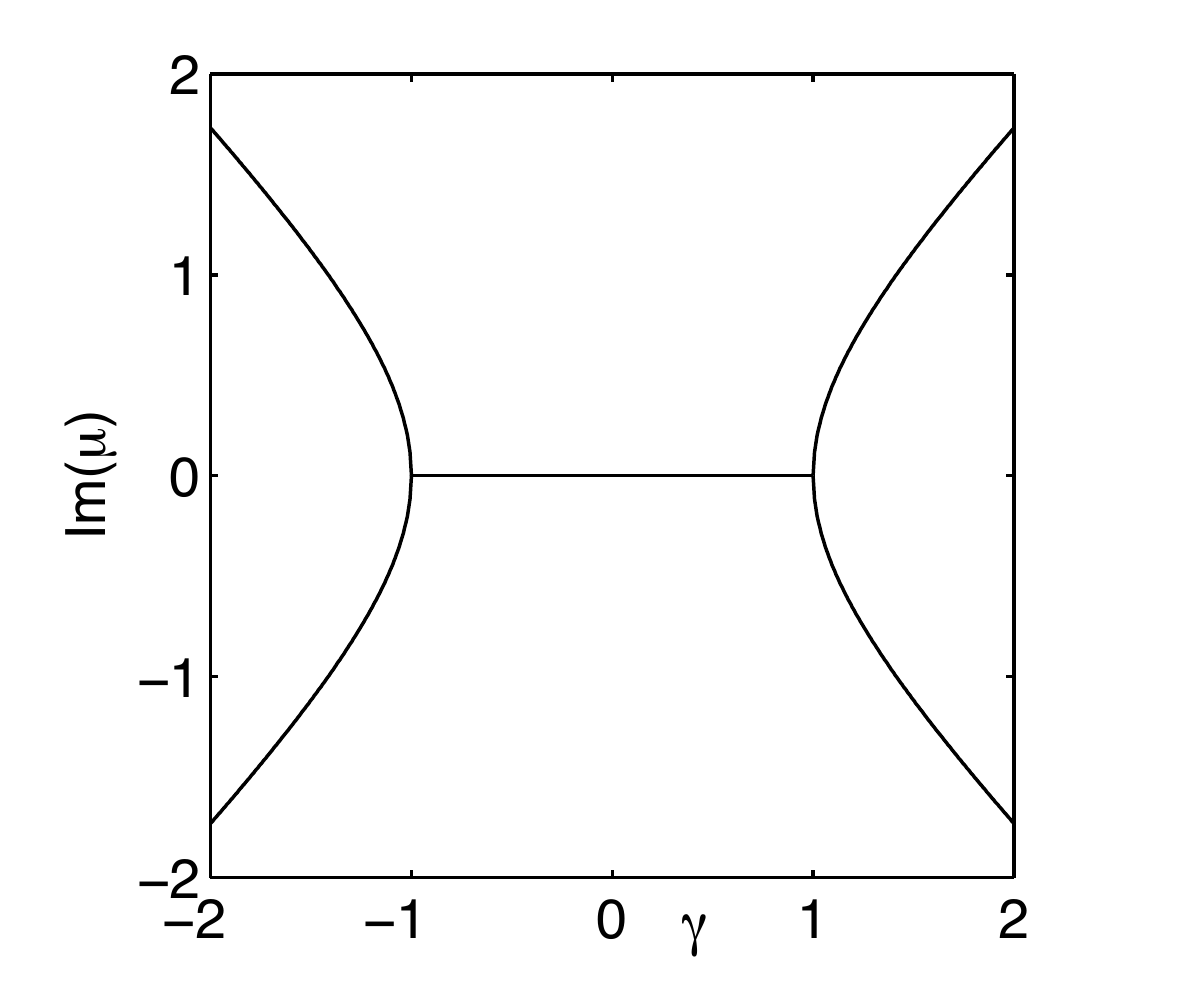}\\
\includegraphics[width=5cm]{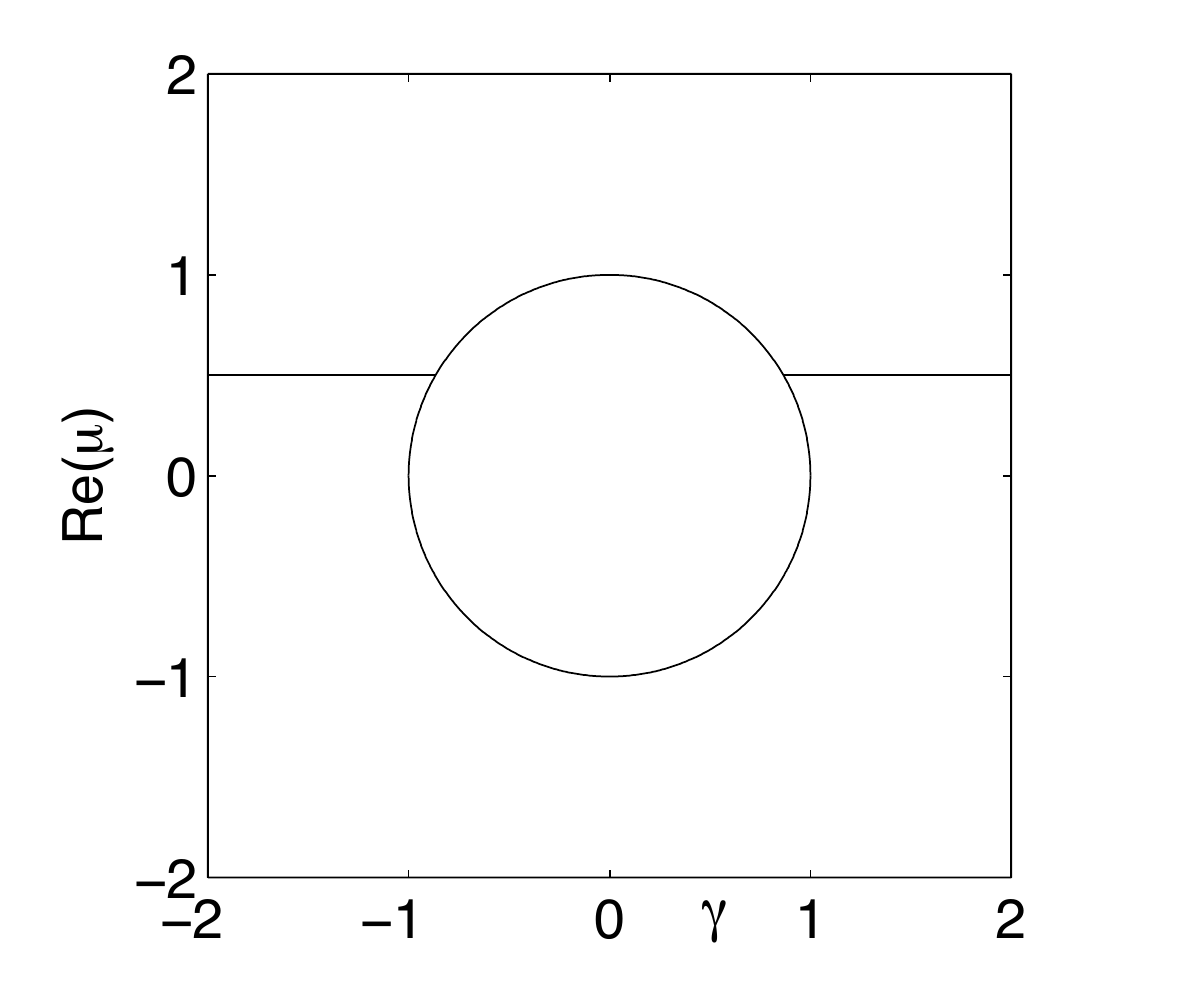}
\includegraphics[width=5cm]{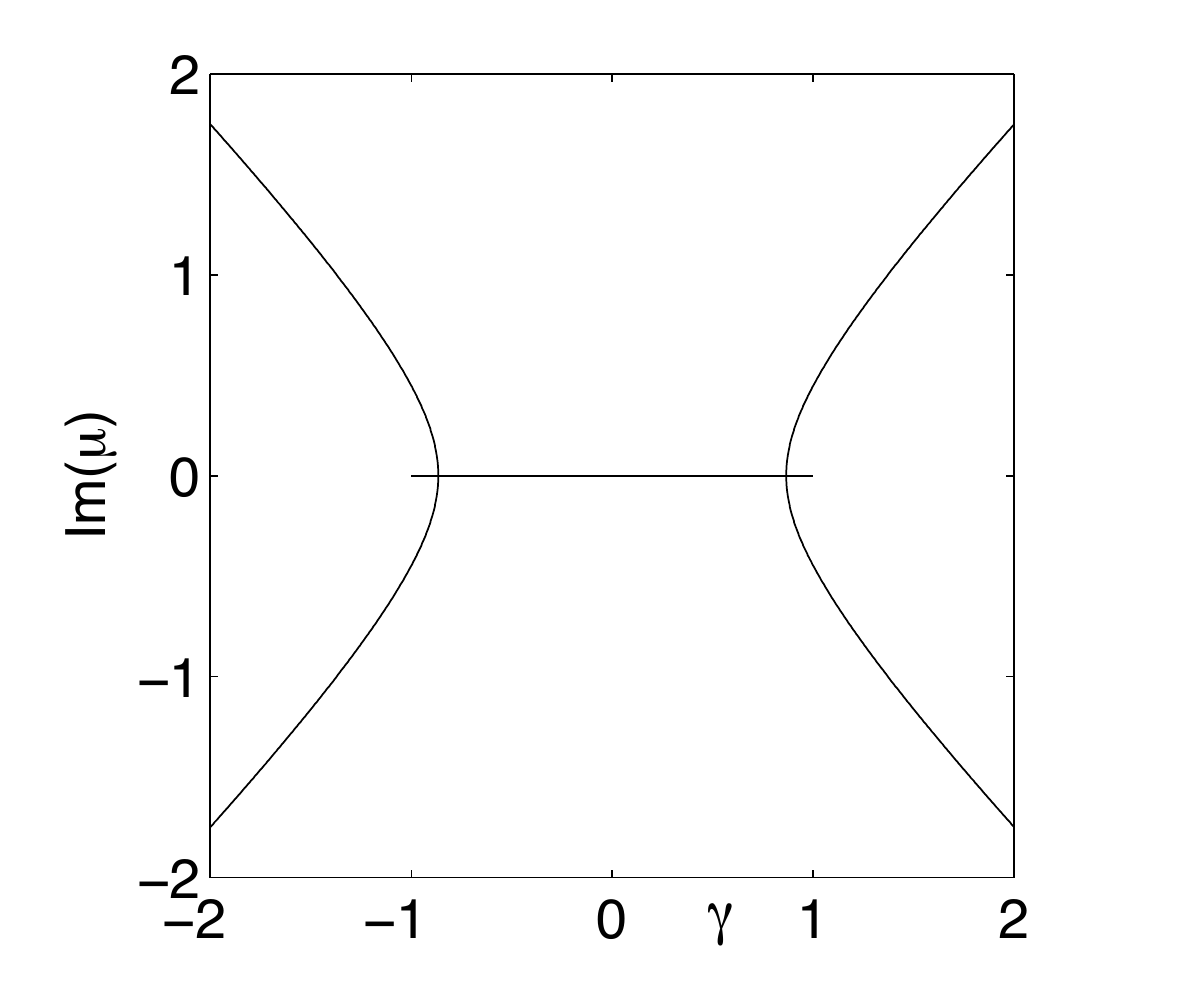}
\end{center}
\caption{\label{fig-2level-nlin-cross-PT}Real (left) and imaginary
(right) part of the complex nonlinear eigenvalues \rf{eqn_mu} as a function of 
$\gamma$ for $v=1$ and $c=0$ 
(top) and $c=0.5$ (bottom).}
\end{figure}

Note that in the nonlinear case the eigenvalues, that is, the  chemical potentials, generally differ from the energy expectation values in the stationary states. The stationary energy values of the system \rf{nlnhGP} have been investigated in \cite{nhbh}.
Closely related eigenvalue problems have been considered in \cite{06nlnh,07nlres,Kevr}. 
We find that for a given stationary solution $\vec{\phi}$
the corresponding eigenvalue can be expressed as
\begin{equation}
\mu=-\rmi\gamma\kappa+v(\phi_1^*\phi_2+\phi_2^*\phi_1)+c\kappa^2.
\label{eqn_mu}
\end{equation}
The energy in this state, on the other hand, is given by 
\begin{equation}
E=-\rmi\gamma\kappa+v(\phi_1^*\phi_2+\phi_2^*\phi_1)+\frac{c}{2}\kappa^2,
\end{equation}
with $\kappa=|\phi_1|^2-|\phi_2|^2$. 
There are up to four stationary solutions, 
which can be seen as follows.
The population imbalance $\kappa=|\phi_1|^2-|\phi_2|^2$ for 
the possible stationary states is given by the real roots 
of the fourth order polynomial \cite{nhbh,06nlnh}
\begin{equation}\label{pol_kappa}
(c^2\!+\!\gamma^2)\kappa^4\!+(v^2
\!-\! c^2\!-\!\gamma^2)\kappa^2\!=\!0,
\end{equation}
which has the four solutions $\kappa=0,\,0,\, \pm \sqrt{1-\frac{v^2}{c^2+\gamma^2 }}$. 
Parameterising the eigenstates as
\begin{equation}
\label{eqn_phi_para}
\phi_1=\sqrt{\frac{\kappa+1}{2}}\,\rme^{-\rmi q},\quad \phi_2=\sqrt{\frac{1-\kappa}{2}}\,\rme^{\rmi q},
\end{equation}
we find from the eigenvalue equation for $\kappa=0$
\begin{equation}
\sin(2q)=\frac{\gamma}{v},
\end{equation}
which leads to two possible values of $q$. For the eigenvalues $\kappa=\pm\sqrt{1-\frac{v^2}{c^2+\gamma^2 }}$, on the other hand, there is only one value of $q$ fulfilling  
\begin{equation}
\cos{(2q)}=\frac{c}{\sqrt{c^2+\gamma^2}}\quad{\rm and}\quad \sin{(2q)}=\frac{\gamma}{\sqrt{c^2+\gamma^2}}.
\end{equation}
In summary, we have the four solutions 
\begin{equation}
\label{eqn_sol_kappa_q}
(\kappa, q)=\left\{\begin{array}{l} \left(0,\, \frac{1}{2}\arcsin\left(\frac{\gamma}{v}\right)\right)\\ \\
\left(\pm \sqrt{1-\frac{v^2}{c^2+\gamma^2 }},\, \frac{1}{2}\arctan\left(\frac{\gamma}{c}\right) \right).\end{array}\right. 
\end{equation} 
However, only those pairs $\kappa, q$ which are real valued correspond to actual solutions of the time independent Schr\"odinger equation \rf{complextigpe}. That is, we have two stationary solutions for $|\gamma|>|v|$ and $\gamma^2<v^2-c^2$, and four stationary states in the remaining region of the parameter space (see also figure 7 in \cite{nhbh}).

Clearly, in the linear case there are two eigenstates for each set of parameters. These are $PT$-symmetric if they are symmetric or antisymmetric with respect to the two basis states, that is, when $\kappa=0$. Thus, $PT$-symmetry is broken at the critical value of the non-Hermitian parameter $|\gamma_{crit} |=|v|$. The two corresponding eigenvalues, depicted in the upper panel in Figure \ref{fig-2level-nlin-cross-PT}, are simply given by $\mu_{\pm}=\pm\sqrt{v^2-\gamma^2}$. They are real in the region of unbroken $PT$ symmetry, and become complex when the eigenstates break the $PT$ symmetry. 

The lower panel of Figure \ref{fig-2level-nlin-cross-PT} shows the nonlinear eigenvalues for a relatively small nonlinearity. We observe a similar behaviour as in the linear case, with two real eigenvalues that coalesce in an exceptional point at the critical value $|\gamma|=|v|$. These eigenvalues are independent of the nonlinearity, as they correspond to $\kappa=0$ and thus the nonlinear term vanishes in the eigenvalue equation \rf{complextigpe}. However, instead of turning to a complex conjugate pair after the exceptional point, the eigenvalues and the corresponding eigenstates vanish in the nonlinear case. On the other hand, there is an additional pair of complex conjugate eigenvalues, which emerge already for smaller values of $|\gamma|$.

\begin{figure}[tb]
\begin{center}
\includegraphics[width=5cm]{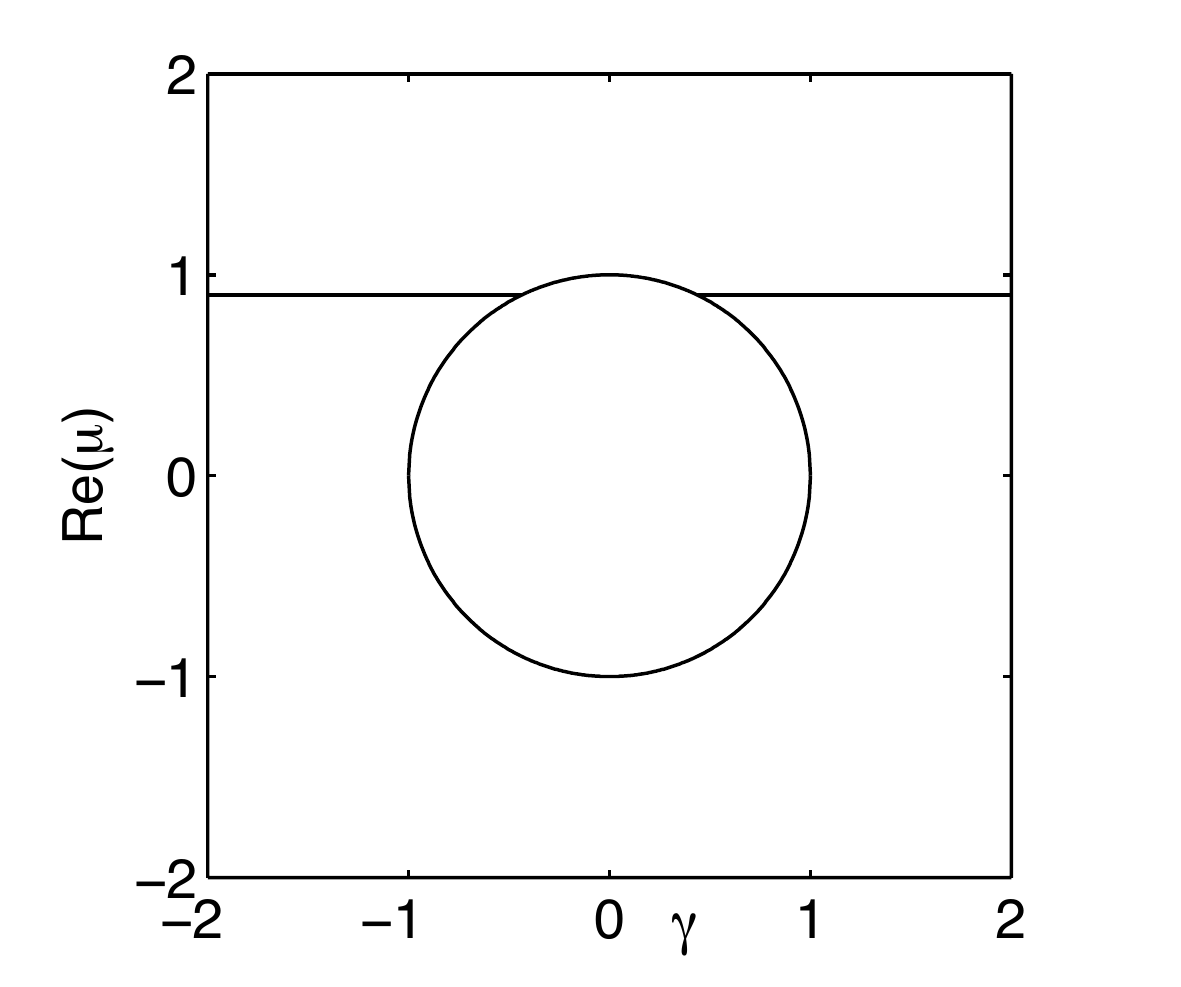}
\includegraphics[width=5cm]{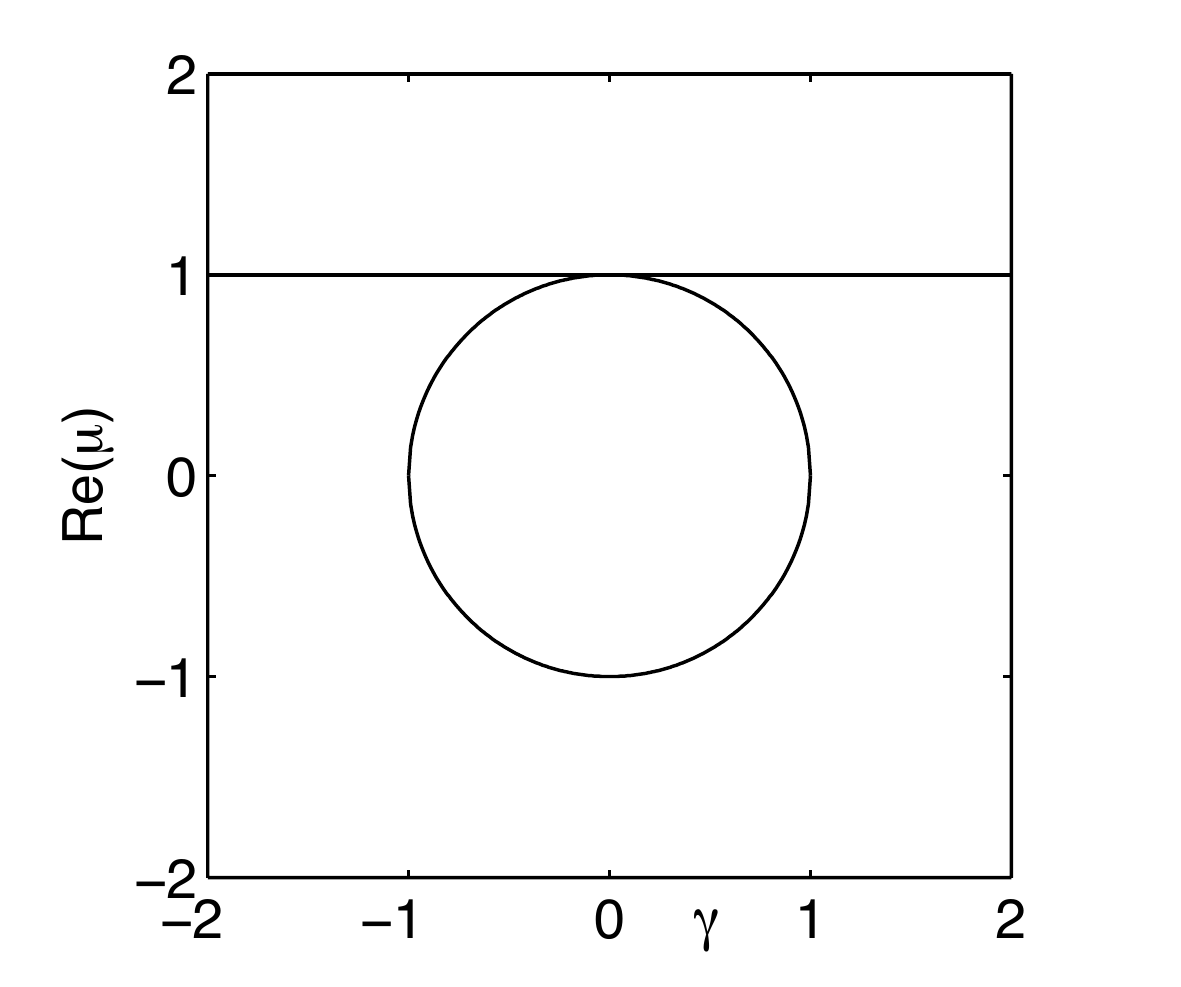}
\includegraphics[width=5cm]{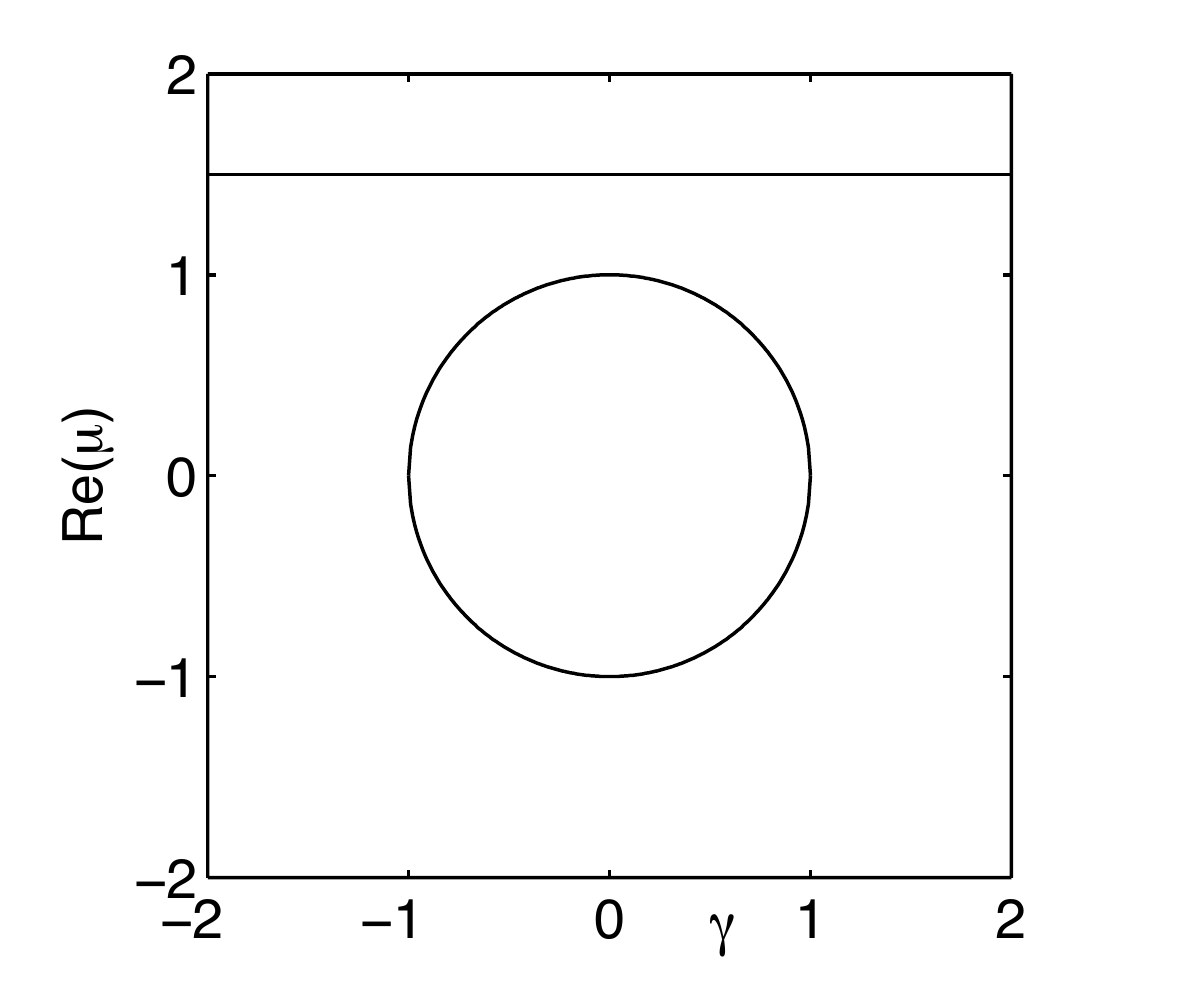}
\includegraphics[width=5cm]{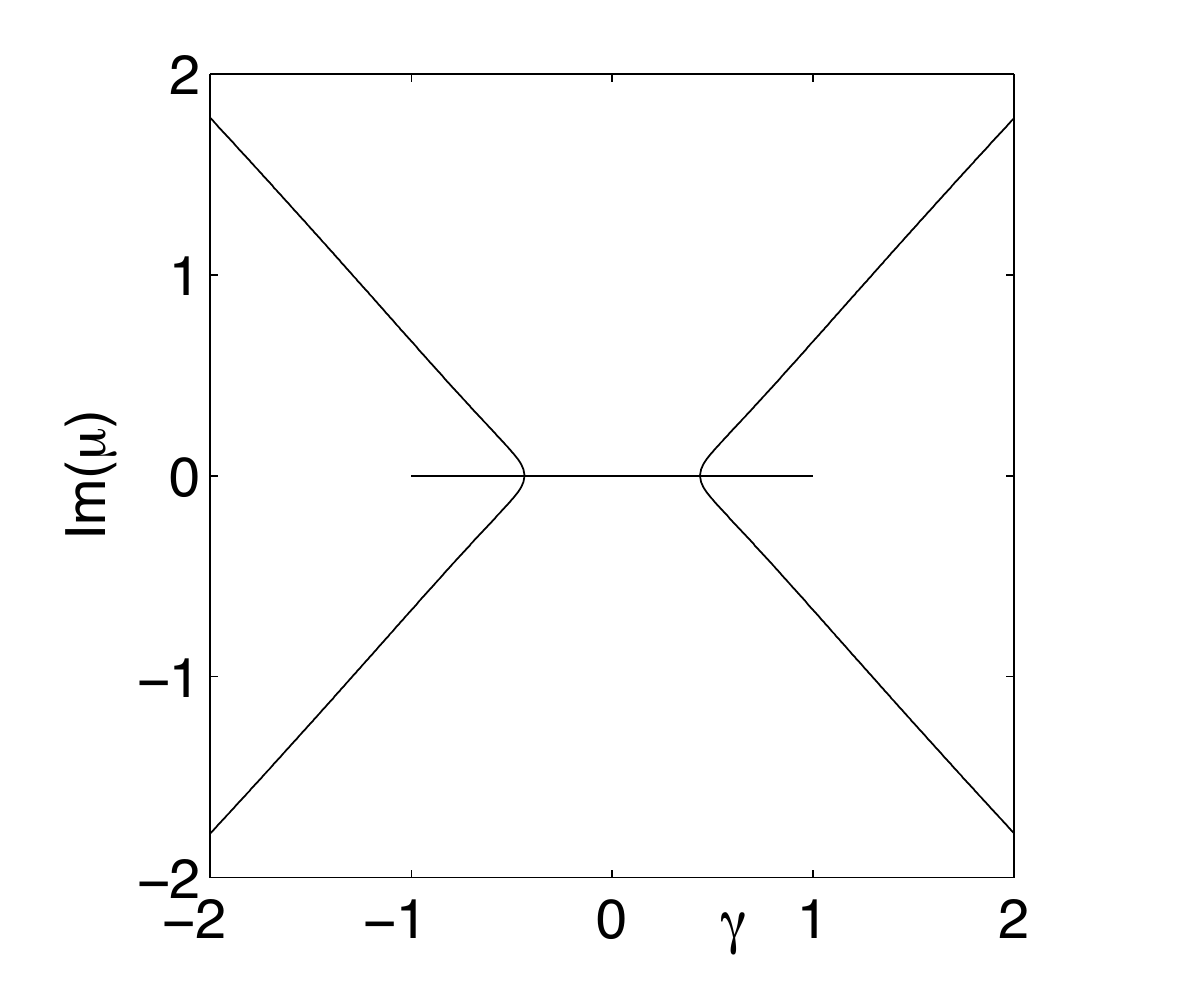}
\includegraphics[width=5cm]{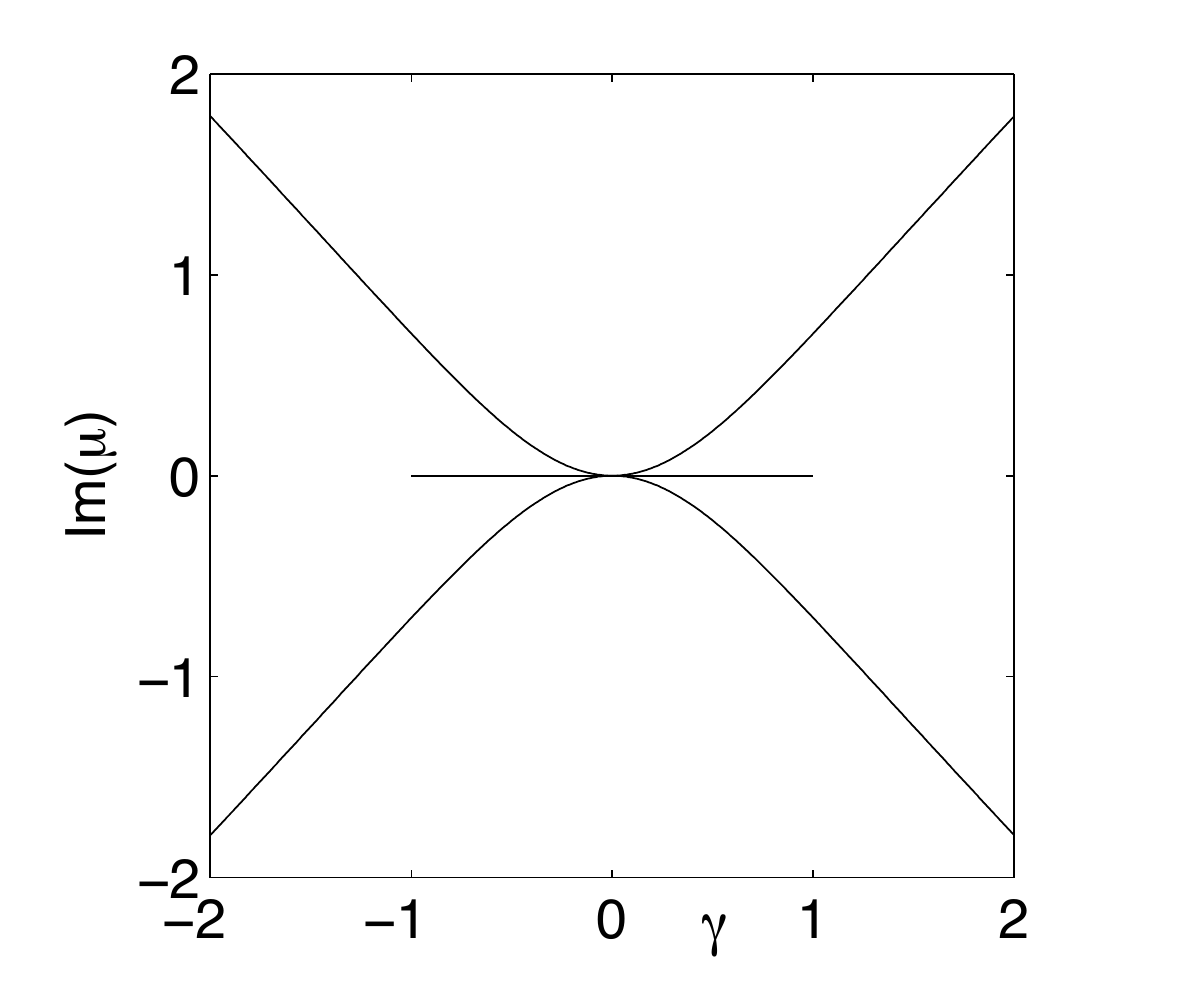}
\includegraphics[width=5cm]{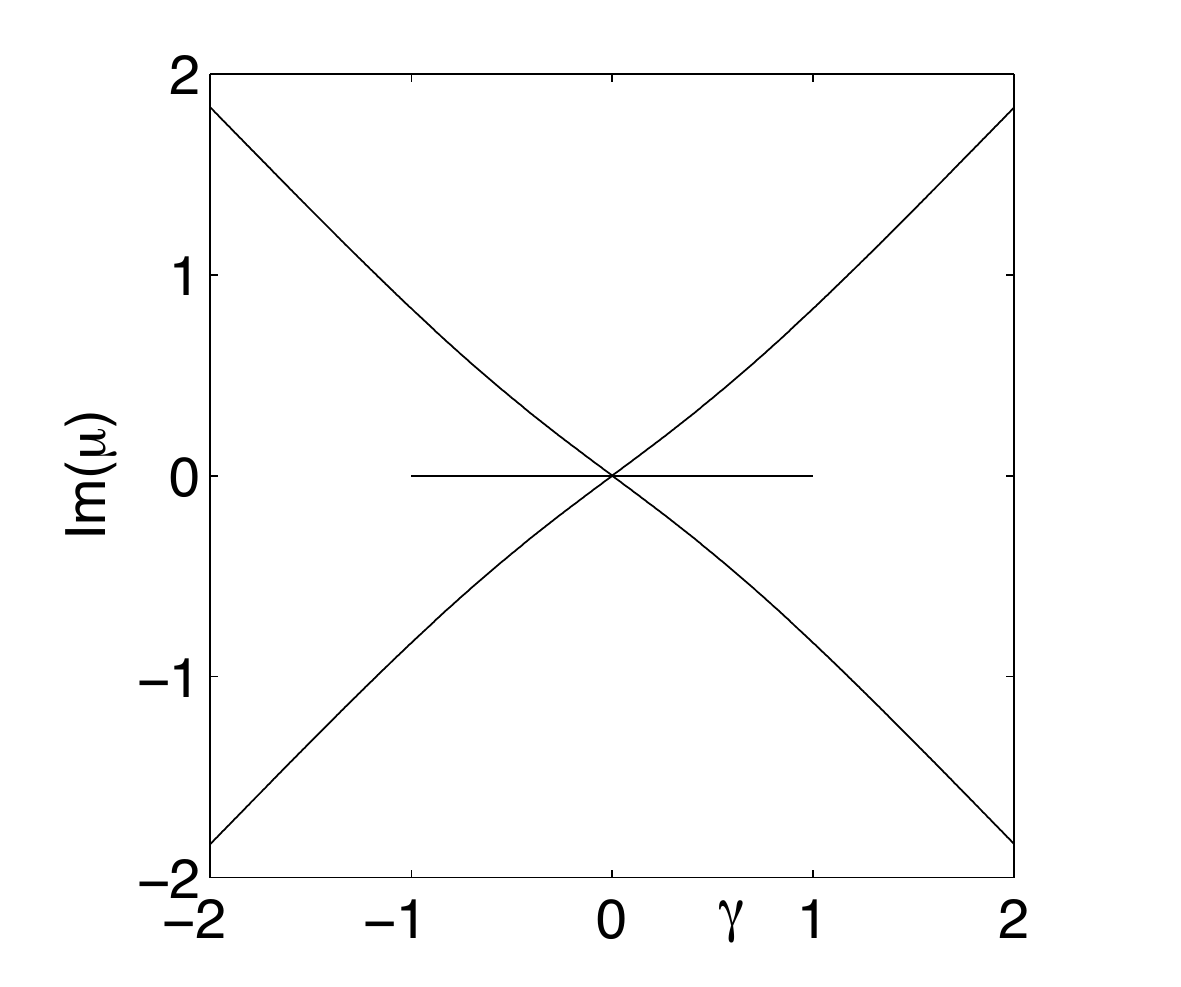}
\end{center}
\caption{\label{fig-2level-nlin-cross-PT2} Real (top) and imaginary
(bottom) part of the complex nonlinear eigenvalues \rf{eqn_mu} as a function of 
$\gamma$ for $v=1$ and $c=0.9$ (left), $c=1$ (middle), and $c=1.5$ (right).}
\end{figure}

In general, in the nonlinear case $c\neq 0$ there can be up to four coexisting eigenstates, only two of which respect the $PT$-symmetry of the system. Figure \ref{fig-2level-nlin-cross-PT2} shows examples of the eigenvalues in dependence on $\gamma$ for different values of $c$. Since the imaginary part 
of the eigenvalue is proportional to the population imbalance $\kappa$, 
$PT$ symmetric solutions correspond to real eigenvalues, as in the linear case. 
The solutions with $\kappa\neq 0$ appear for values of the non-Hermiticity parameter for which $\gamma^2\geq v^2-c^2$. Thus, in the presence of nonlinearity the $PT$-symmetry of the system is broken for smaller values of the non-Hermiticity, and above the critical value $|c_{crit}|=|v|$ there is no region of unbroken $PT$ symmetry at all. 

In the Hermitian case additional states induced by the nonlinearity are often referred to as self-trapping states, because they favour one of the modes, thus breaking the symmetry of the linear part of the system. Therefore, one can argue that the non-Hermiticity enhances the self trapping effect, by reducing the critical nonlinearity for which additional states emerge, however, on cost of the $PT$-symmetry of the system. 

It is well known that the self-trapping transition in the two-mode Gross-Pitaevskii equation with real parameters is accompanied by a restructuring of the dynamical stability of the eigenstates: While the self-trapping modes that break the symmetry of the system are dynamically stable, the previously stable antisymmetric state becomes dynamically unstable after the self-trapping transition. The stability can be determined via the Bogoliubov-de-Gennes equation for the stationary states, which arise from the linearised effect of a small time-dependent perturbation of the stationary solution of the form $\vec{\psi}(t)=\rme^{-i\mu t}\left(\vec{\phi}+\vec{\delta\phi_{-}}\rme^{-\rmi\omega t}+\vec{\delta\phi_{+}^*}\rme^{\rmi\omega^* t}\right)$. However, in the case of complex chemical potential, care has to be taken to consider perturbations that are orthogonal to the stationary solution $\vec\phi$ \cite{Cast98,09ddshell}. In the present case the resulting Bogoliubov-de-Gennes equations reduce to the eigenvalue problem 
\begin{equation}
B\left(\!\!\begin{array}{c} \vec{\delta\phi_{-}}\\ \vec{\delta\phi_{+}} \\ \end{array}\!\!\right)=\omega \left(\!\!\begin{array}{c} \vec{\delta\phi_{-}}\\ \vec{\delta\phi_{+}} \\ \end{array}\!\!\right),
\end{equation}
with the $4\times4$ matrix 
\begin{equation}
\nonumber
B\!=\!\left(\!\!\!\begin{array}{cc}
 H_{0\,nl}\!-\!\mu\!+\!c\, Q\!\left(\!\begin{array}{cc}|\phi_1|^2& -\phi_1\phi_2^*\\ -\phi_1^*\phi_2& |\phi_2|^2\end{array}\!\right)\!Q & c\, Q\!\left(\!\begin{array}{cc}\phi_1^2& -\phi_1\phi_2\\ -\phi_1\phi_2& \phi_2^2\end{array}\!\right)\! Q^*\\
-c\, Q^*\! \left(\!\begin{array}{cc}\phi_1^{*2}& -\phi_1^*\phi_2^*\\ -\phi_1^*\phi_2^*& \phi_2^{*2}\end{array}\!\right)\! Q &
-H_{0\, nl}^*\!+\!\mu^*\!-\!c\, Q^*\!\left(\!\begin{array}{cc}|\phi_1|^2& -\phi_1^*\phi_2\\ -\phi_1\phi_2^*& |\phi_2|^2\end{array}\!\right)\!Q^*
\end{array}\!\!\!\right).
\end{equation}
Here $Q$ denotes the projection operator orthogonal to the stationary solution $\vec{\phi}$, 
$$Q=\begin{pmatrix} 1-|\phi_1|^2 & -\phi_1\phi_2^* \\
  -\phi_1^*\phi_2 &  1 -|\phi_2|^2 \end{pmatrix}$$ 
and $H_{0\, nl}$ is the complex nonlinear Hamiltonian evaluated at the stationary solution
$$ H_{0\, nl}=\begin{pmatrix} - \rmi \gamma + c( |\phi_1|^2-|\phi_2|^2)  & v \\
  v &  \rmi \gamma -c (|\phi_1|^2 -|\phi_2|^2) \end{pmatrix}. $$ 
The stability of a stationary solution is determined by the imaginary part of the eigenvalues $\omega$: While for $\Im(\omega)\leq0$ the perturbations are decaying or purely oscillating, they grow exponentially for $\Im(\omega)>0$, in which case the stationary solution is dynamically unstable. 

\begin{figure}[tb]
\begin{center}
\includegraphics[width=5.1cm]{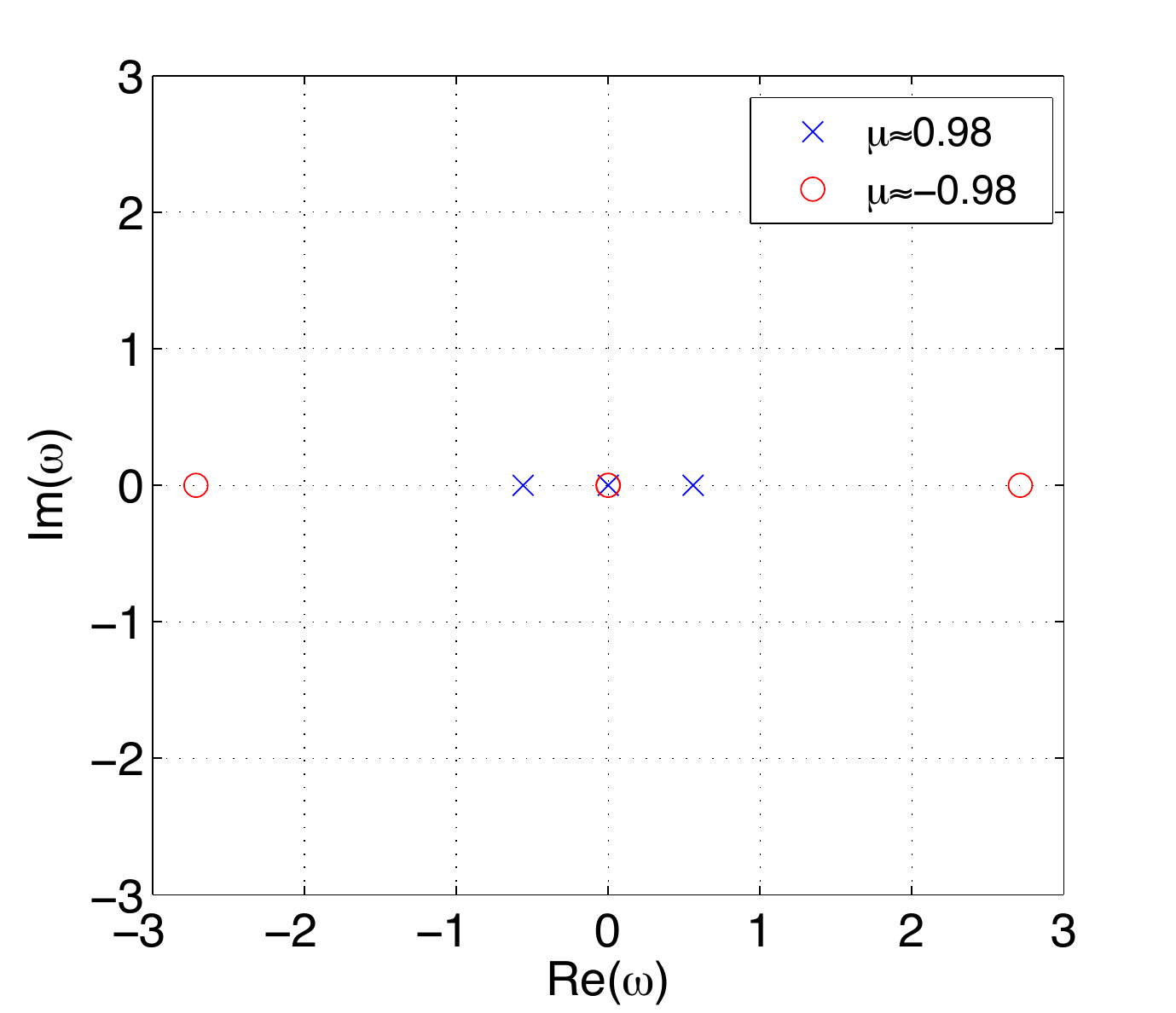}
\includegraphics[width=5.1cm]{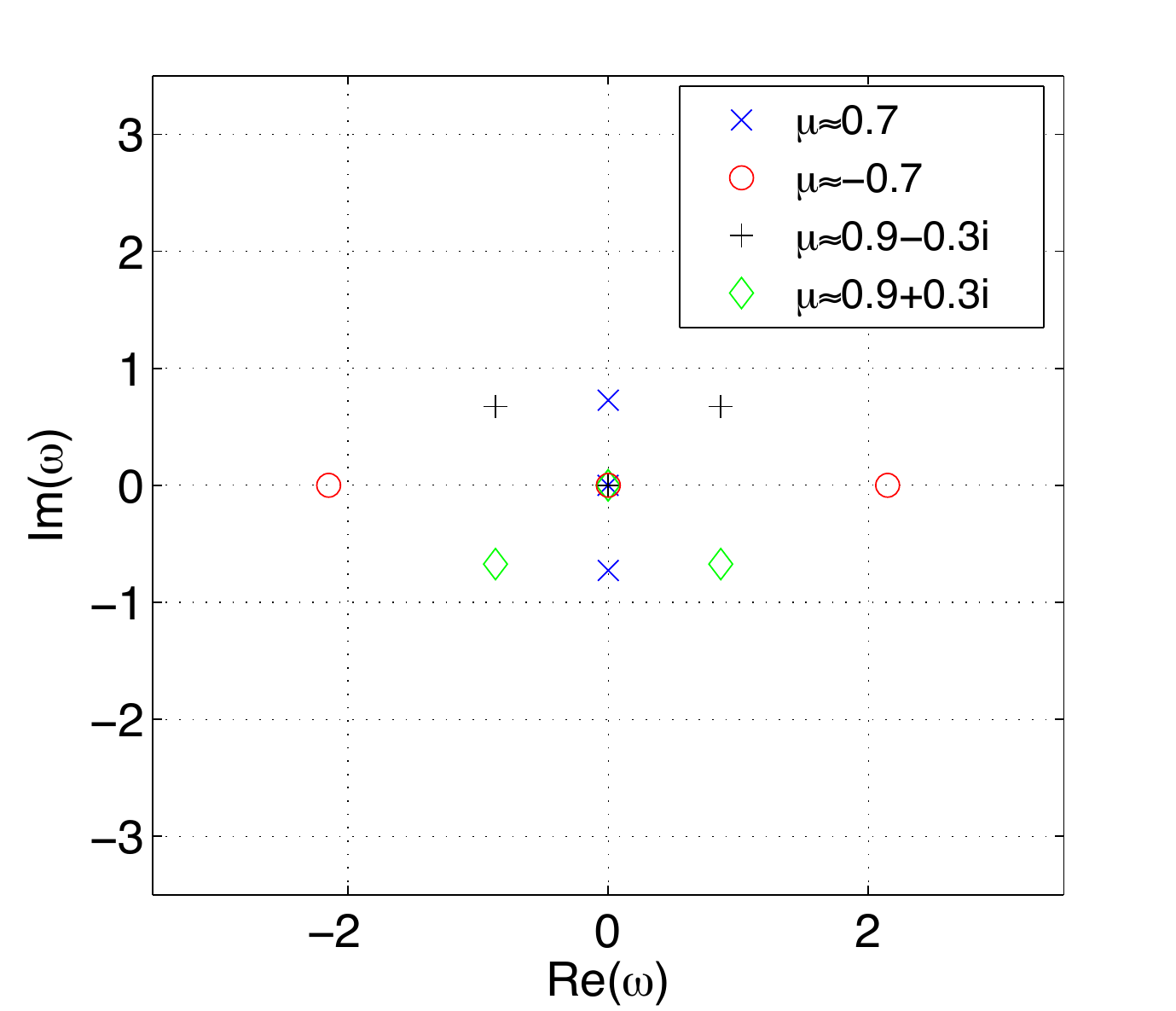}
\includegraphics[width=5.1cm]{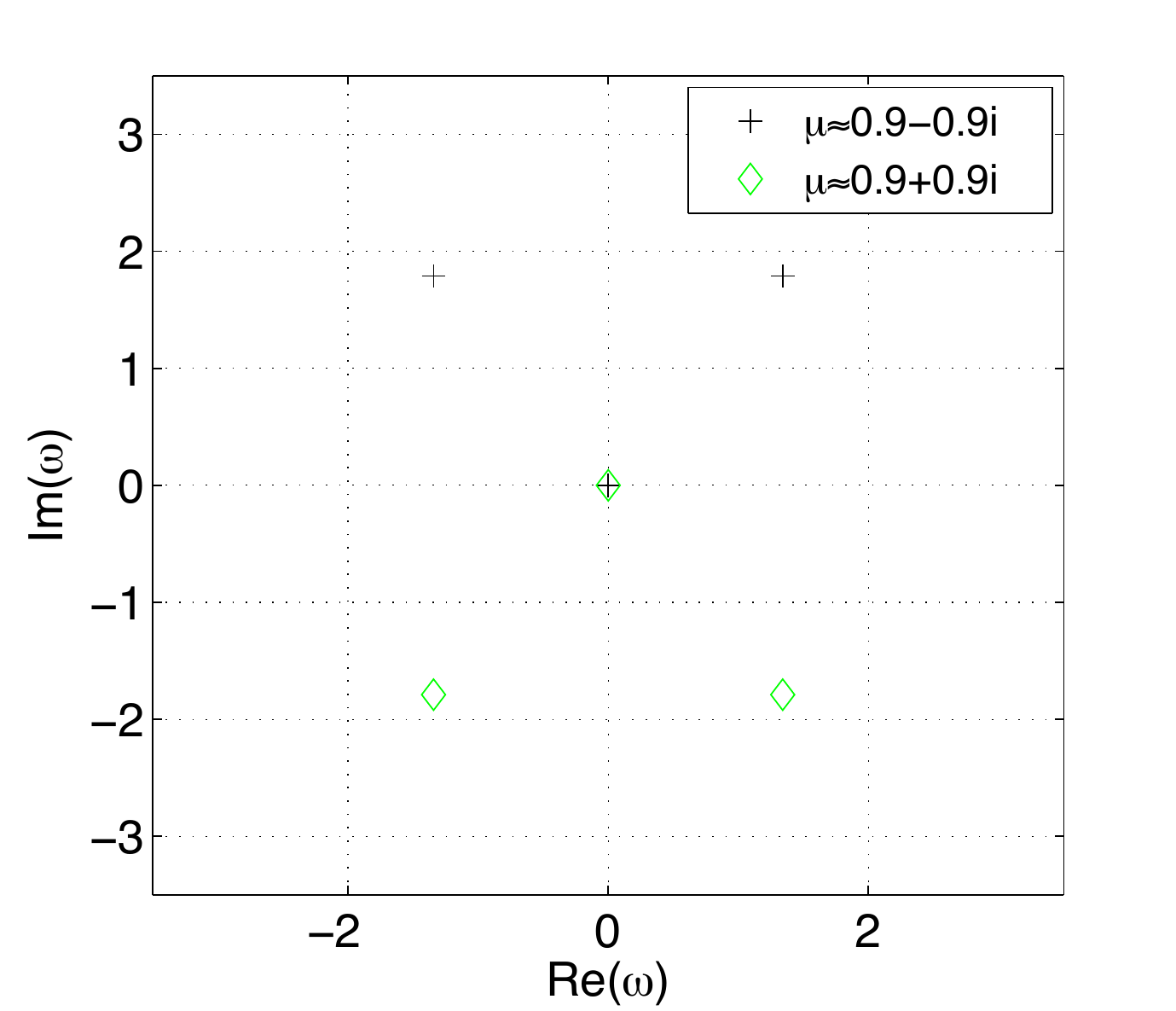}
\end{center}
\caption{\label{fig-stability} Eigenvalues $\omega$ of the stability matrix $B$ corresponding to the different stationary states in the complex plane for $c=0.9$, $v=1$ and $\gamma=0.2$ (left), $\gamma=0.7$ (middle), and $\gamma=1.2$ (right).}
\end{figure}

Figure \ref{fig-stability} shows the numerically obtained eigenvalues $\omega$ in the complex plane for all stationary states for $c=0.5$, $v=1$ and three different values of $\gamma$. The figure in the left panel corresponds to $\gamma=0.2$, where there are only two eigenstates with real chemical potentials, which are both stable. For $\gamma=0.7$, already in the region of broken $PT$-symmetry, there are two additional eigenstates with complex conjugate chemical potentials. The one with a positive imaginary part of the chemical potential is dynamically stable, while the other one is unstable. This is related to the fact that these states appear as local sink and source of the dynamics \cite{08nhbh_s,nhbh}. As in the Hermitian case, one of the remaining stationary states is now unstable, it is a saddle point of the dynamics \cite{08nhbh_s,nhbh}. It can be easily verified analytically that this state changes its dynamical behaviour at the point in parameter space at which the additional eigenstates emerge: The eigenvalue equation for the stability matrix $B$ for the states $(\kappa,q)=\left(0,\, \frac{1}{2}\arcsin\left(\frac{\gamma}{v}\right)\right)$, that correspond to the chemical potentials $\mu=\pm\sqrt{v^2-\gamma^2}$ reduces to
\begin{equation}
\omega^4+4(\gamma^2-v^2\pm c\sqrt{v^2-\gamma^2})\omega^2=0.
\end{equation}
Thus, for positive $c$ the upper state becomes dynamically unstable for $\gamma>\sqrt{v^2-c^2}$, while the other state stays stable. The situation is reversed for negative values of $c$, in this latter case it is also the lower chemical potential from which the additional eigenvalues branch off. This behaviour is in complete analogy to the self-trapping transition for $\gamma=0$.
For even larger values of $\gamma>1$ only the two eigenstates with complex chemical potential remain, which do not change their stability, that is one of them is stable, the other is unstable.

We have seen that the complex nonlinear eigenvalue problem \rf{complextigpe} leads to a varying number of solutions, depending on the precise parameter values, where eigenvalues may appear, vanish and change their stability when parameters are varied. While such bifurcations are typical in nonlinear problems, it might lead to useful insights to consider an extension of \rf{complextigpe} that has four eigenvalues for all parameter values, which coincide with the eigenvalues of the original system wherever the latter are defined. We will introduce such an extension in the following section.
 
\section{Analytic extension of the Gross-Pitaevskii equation and an isospectral linear system} 
\label{sec_iso}
\begin{figure}[tb]
\begin{center}
\includegraphics[width=5cm]{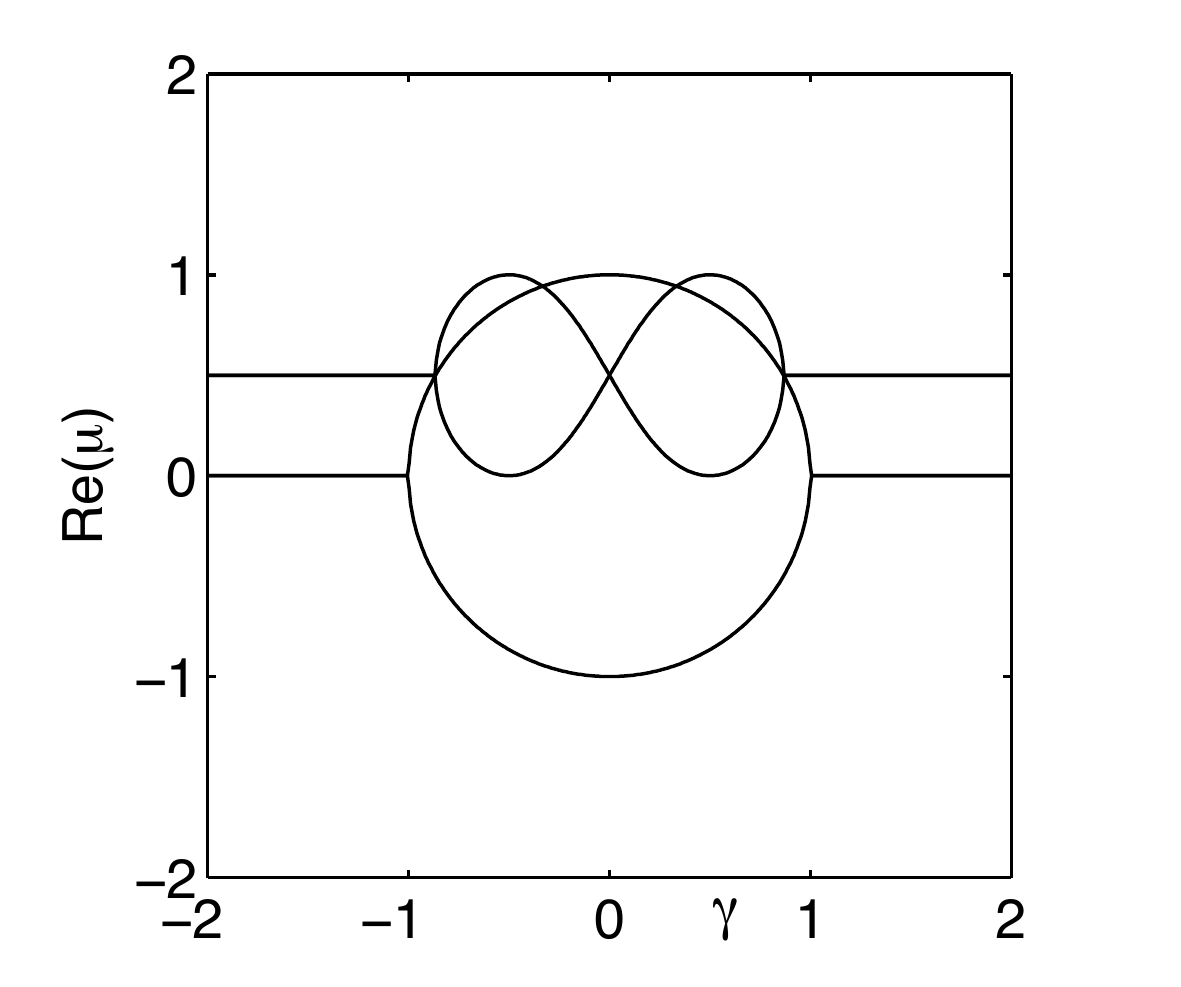}
\includegraphics[width=5cm]{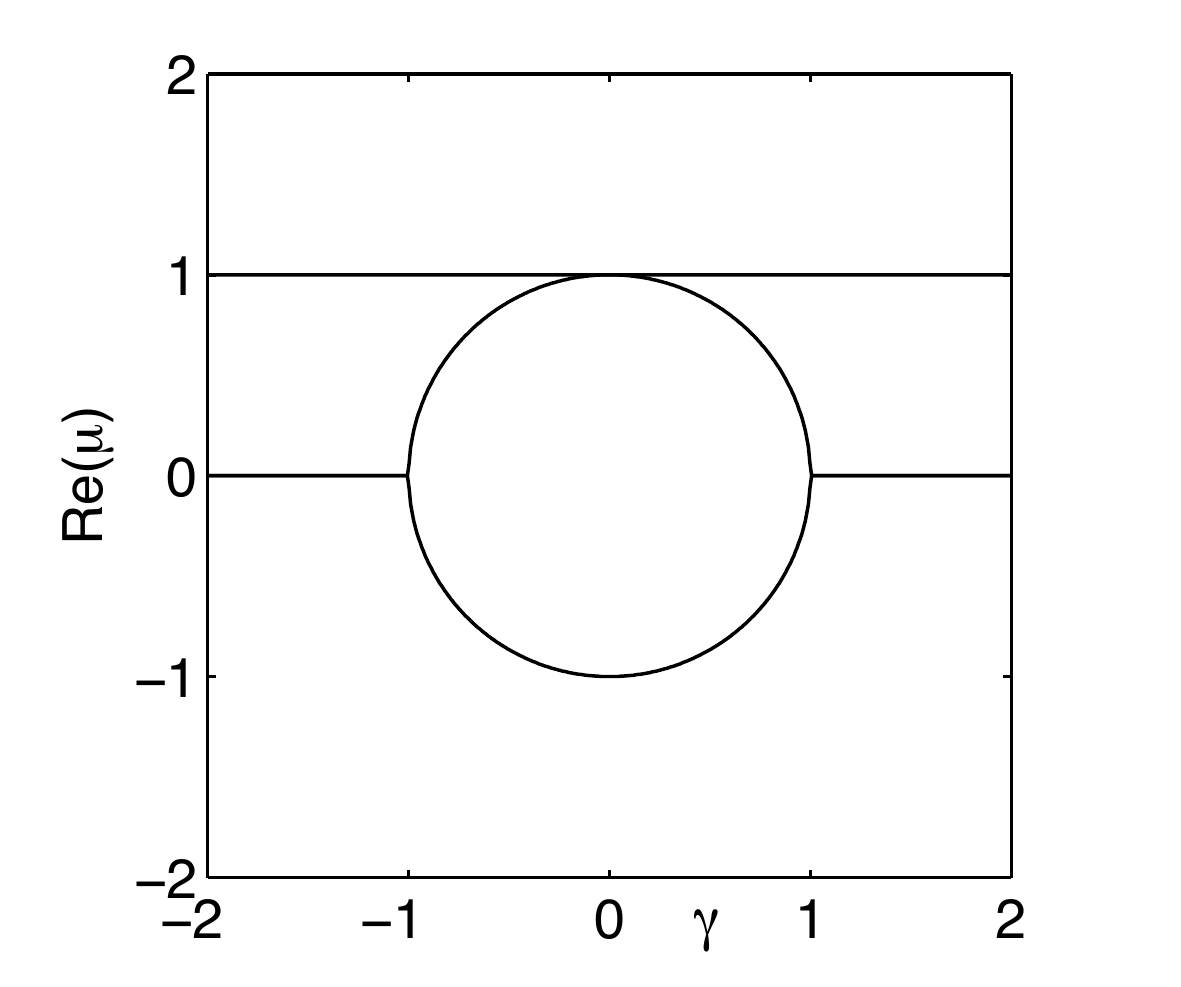}
\includegraphics[width=5cm]{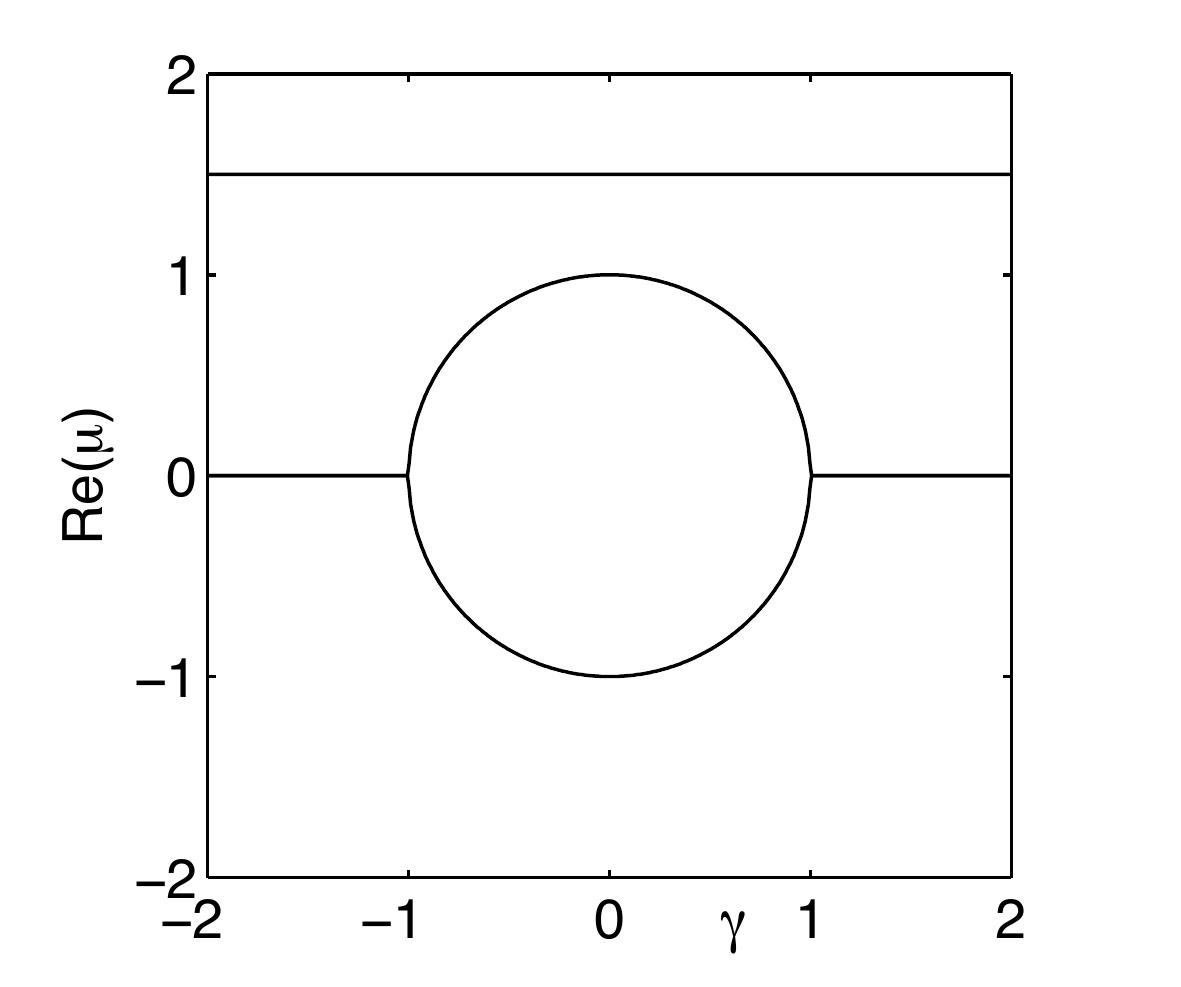}
\includegraphics[width=5cm]{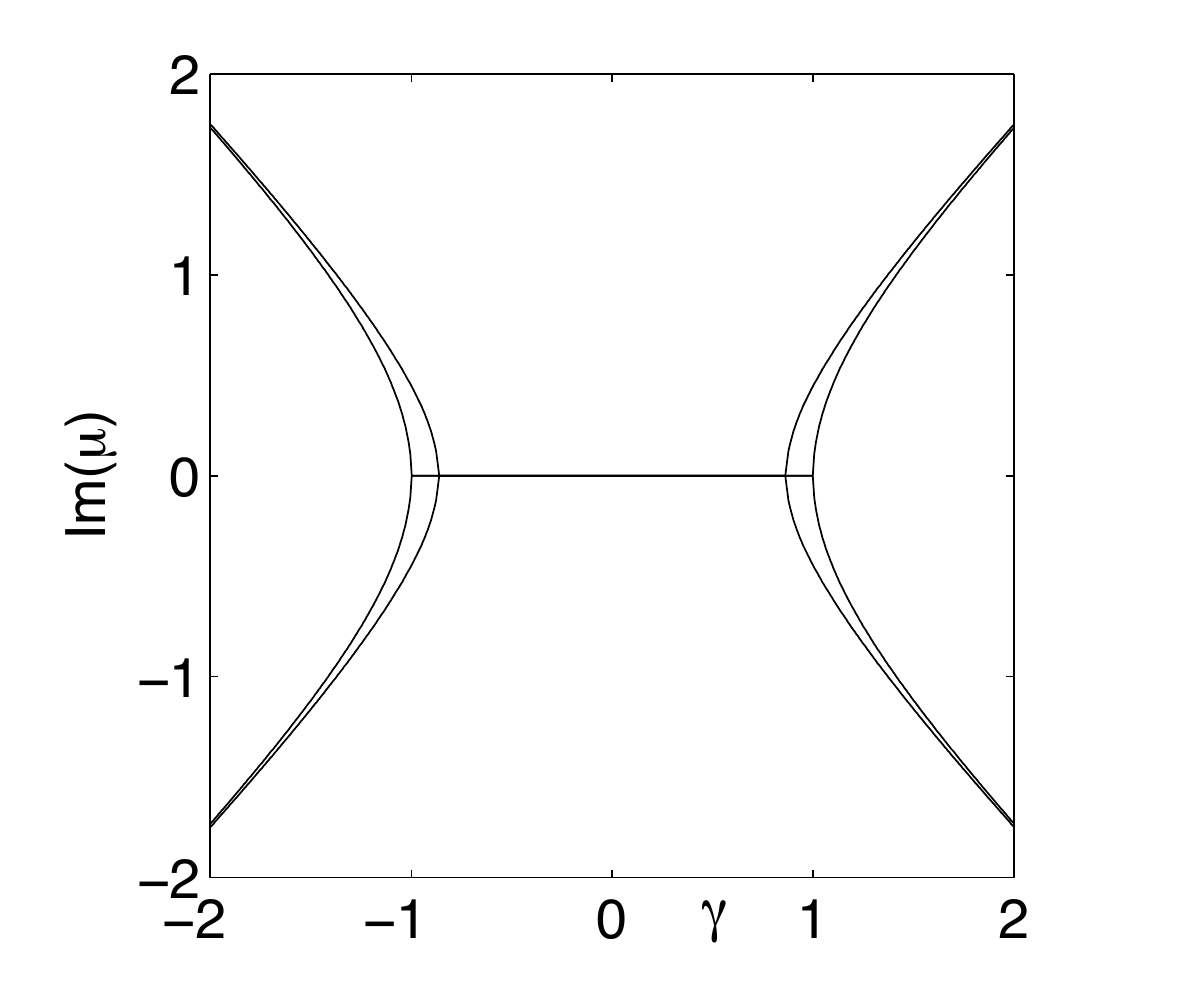}
\includegraphics[width=5cm]{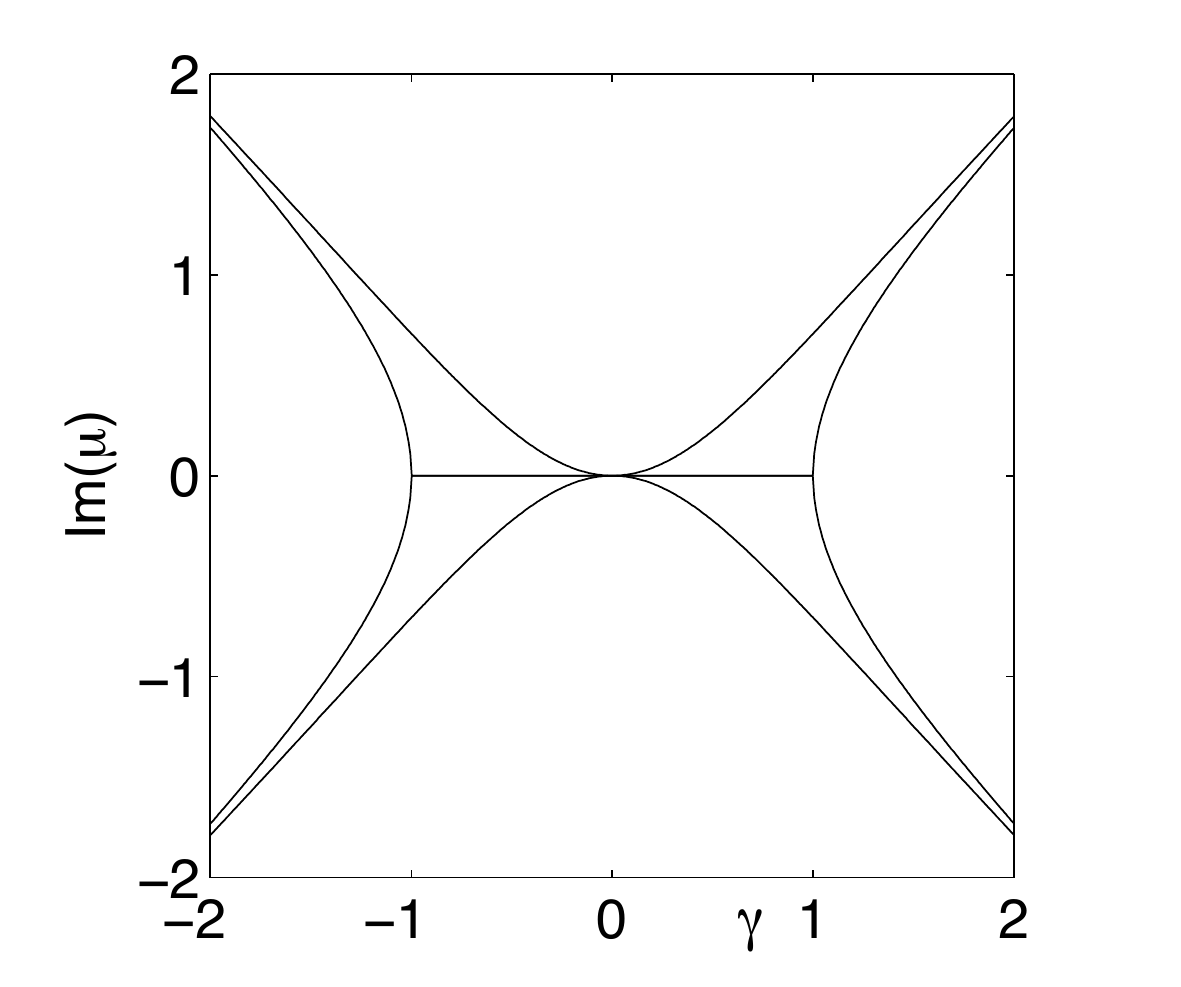}
\includegraphics[width=5cm]{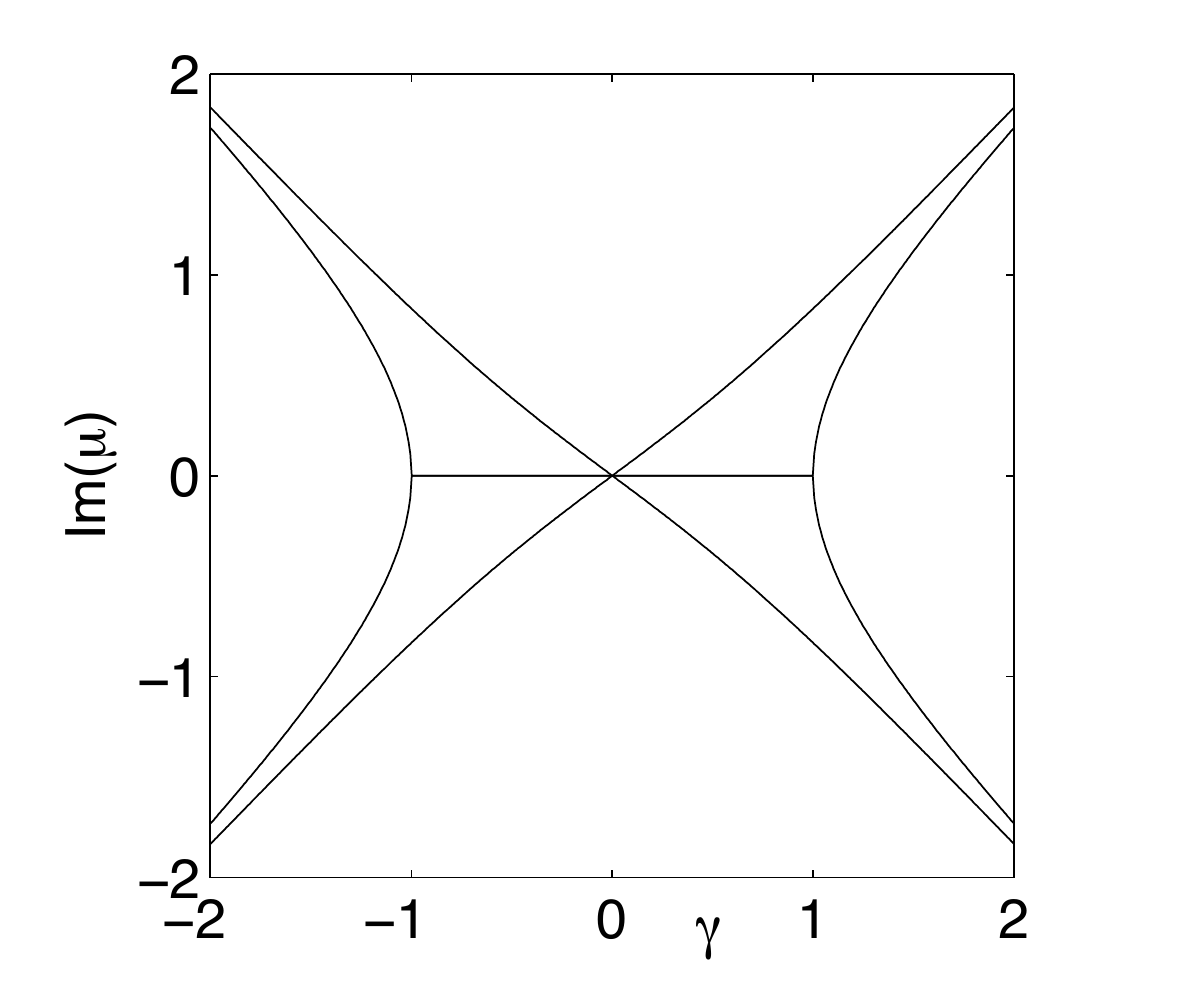}
\end{center}
\caption{\label{fig-4level-cross-PT} Real (top) and imaginary
(bottom) part of the complex continued mean-field eigenvalues \rf{eqn_mu_cont} as a function of $\gamma$ for $v=1$ and $c=0.5$ (left), $c=1$ (center) and $c=1.5$ (right).}
\end{figure}

The vanishing of solutions in the Gross-Pitaevskii equation is related to the fact that the nonlinear term $g|\psi|^2$ is not analytic \cite{Cart08,Wunn12}, and the ``missing'' eigenstates and eigenvalues can be recovered by considering analytic extensions of the problem. The idea presented in \cite{Cart08} is as follows: Every wave function $\psi(x)$ can formally be decomposed in terms of two real valued functions $A(x)$ and $\varphi(x)$ corresponding to the amplitude and the phase of $\psi(x)=A(x)\rme^{\rmi \varphi(x)}$. The non-analytic complex Gross-Pitaevskii equation can thus be decomposed into two coupled real differential equations. To extend the parameters into the complex domain, rather than considering the original equation, one formally considers the equations for $A(x)$ and $\varphi(x)$, which now lead to complex solutions. This is equivalent to replacing $|\psi(x)|^2$ in the Gross-Pitaevskii equation with the complex valued function $A^2(x)$. 
In the present case, this leads to the complex Gross-Pitaevskii equation
\begin{equation}
\label{complextigpe_cont}
\begin{pmatrix} - \rmi \gamma + c \kappa & v \\
  v &  \rmi \gamma - c\kappa \end{pmatrix} \left(\begin{array}{c} \phi_1 \\ \phi_2 \\ \end{array}\right)=\mu \left(\begin{array}{c} \phi_1 \\ \phi_2 \\ \end{array}\right),
 \end{equation}
where the eigenstates $(\phi_1,\phi_2)$ are parameterised according to \rf{eqn_phi_para}, with arbitrary $\kappa,q\in\mathds{C}$. When solving the eigenvalue equation we have in fact already encountered the additional solutions, which led to complex values of $\kappa$ or $q$ in \rf{eqn_sol_kappa_q}. 
The corresponding eigenvalues are then given by
\begin{equation}
\mu=(-\rmi\gamma+c\kappa)\kappa+v\sqrt{1-\kappa^2}\cos(2q), 
\label{eqn_mu_cont}
\end{equation}
for all solutions $(\kappa,q)$ in \rf{eqn_sol_kappa_q}. 
Substituting the values of $\kappa$ and $q$ into \rf{eqn_mu_cont} we find the eigenvalues
\begin{equation}
\mu=\left\{\begin{array}{l}\pm\sqrt{v^2-\gamma^2}\\
c\pm\gamma\sqrt{\frac{v^2}{c^2+\gamma^2}-1}\, , \end{array}\right.
\end{equation}
that is, the eigenvalues of the linear problem and two additional nonlinear eigenvalues. Thus, the eigenvalues are naturally divided into two separate sets.
The resulting spectrum, as a function of $\gamma$, is depicted in Figure \ref{fig-4level-cross-PT}. In the limit $c\to0$ the two nonlinear eigenvalues approach the linear ones. It should be noted that in this limit also the nonlinear eigenstates coincide with the linear ones, and thus we recover two copies of the linear system. 

It is an interesting question whether spectral features of nonlinear systems can be mimicked by suitable linear systems, and which symmetries these linear systems exhibit. In the present case the spectrum of the analytically continued nonlinear eigenvalue problem \rf{complextigpe_cont} can be immediately interpreted as that of an equivalent four-dimensional linear system. Since the eigenvalues decouple into two sets, it is not difficult to construct a matrix representation of this linear system. It is straightforward to verify that the nonlinear problem \rf{complextigpe_cont} is isospectral to the linear $4\times 4$ matrix:
\begin{equation}
\label{eqn_equ_matrix}
M=\left(\!\!\begin{array}{cccc}
 -\rmi\gamma & \ \ v& 0& 0\\
\ \ v& \ \ \rmi\gamma  & 0 & 0\\
\ \ 0 & \ \ 0 & -\rmi\gamma+c & \frac{\rmi\gamma v}{c+\rmi\gamma}\\
\ \ 0 & \ \ 0 & \frac{-\rmi\gamma v}{c-\rmi\gamma} &\rmi\gamma+c
\end{array}\!\!\right).
\end{equation}
This matrix is itself $PT$-symmetric. However, there is an interesting subtlety to this: While the upper block has the same symmetry as the linear two-level system, the lower block has additional complex elements on the off-diagonal, and is not $PT$ symmetric with respect to the standard $T$ operator defined as complex conjugation alone. Instead, the second block is $PT$ symmetric with respect to the $P$ operator exchanging the two relevant basis elements, and the $T$ operator defined as Hermitian conjugation. 

Let us finally turn to the relation between the eigenstates of the two systems. On a formal level, they are in fact directly related: The eigenstates of the nonlinear two-level system \rf{complextigpe_cont} appear in the upper and lower components of the eigenvectors of the matrix \rf{eqn_equ_matrix}, while the remaining components automatically vanish due to the block diagonal form of the system. This, however means, that there is no coupling or overlap between the two sets of eigenstates, and in particular the degeneracy between one of the eigenvalues of the upper block with the degenerate pair of eigenvalues of the lower block at $\gamma=\sqrt{v^2-c^2}$ has to be interpreted as ocuring incidentally in the linear system. Since the dynamics of the two systems are obviously different, it is not surprising that also the dynamical stability properties of the eigenstates differ. In the linear model it can be checked in a straightforward manner that all eigenstates are stable in the region of unbroken $PT$-symmetry. This includes the Hermitian limit $\gamma\to0$, which means that the reorganisation of the stability properties due to the self-trapping transition is not reflected in the linear system. On the other hand, once the $PT$-symmetry is broken, only the eigenstate corresponding to the eigenvalue with the largest imaginary part is dynamically stable in the linear system, leaving the remaining three states unstable. This is connected to the fact that the eigenstate corresponding to the eigenvalue with the largest imaginary part acts as a global sink of the dynamics. Thus, the equivalence between the linear and the original nonlinear complex eigenvalue problem clearly has limitations, in particular in regard to the related dynamical properties. It is, however, an interesting question whether there might be a hidden deeper relation between the models, for example in a sense of constraint quantum mechanics \cite{constrQM}. 

The final interpretation of the isospectral linear operator for the nonlinear eigenvalue problem is an open problem, the relevance of which will only become apparent in the study of more general systems. A deeper understanding of the relation between the two systems might lead to valuable insights into the nature of both real and complex nonlinear eigenvalue problems, and the nature of complex symmetries. The investigation of these issues, however, goes beyond the scope of the present paper. 

\section{Summary and Outlook}
\label{sec_sum}
We have presented an analytically solvable paradigm system for complexified nonlinear quantum systems. The model can be interpreted as a simple description of a BEC in a $PT$ symmetric double well potential. In particular, we have identified and analysed the eigenvalues and corresponding eigenstates of the system and their stability properties. Similar to the Hermitian case, the nonlinearity can lead to additional eigenstates, which in the complexified case leads to a breaking of $PT$-symmetry. This analytically solvable model is expected to be valuable in the understanding of more complicated systems, which can only be investigated numerically. An example is the model of a BEC in a non-Hermitian double delta potential, numerically investigated in \cite{Wunn12}, where similar features to those discussed here have been found. 

We have further introduced a continuation of the problem that is isospectral to a simple matrix model, whose eigenstates are also related to those of the nonlinear system. In the present case this system was directly constructed from the known solutions of the nonlinear system. If a more general connection between these isospectral models could be uncovered this would shed new light onto the interpretation of nonlinear eigenvalue problems. To gain further insights into this connection, it would be useful to investigate more general systems, for example, discrete nonlinear Schr\"odinger equations with additional modes or systems with different types of nonlinearities.

\vskip 9pt
I thank G\"unter Wunner for stimulating discussions, and Hugh Jones, Dorje Brody, and Hans J\"urgen Korsch for useful comments. I would like to express my gratitude to Nimrod Moiseyev and the Chemistry Department of the Technion, Haifa for kind hospitality while this manuscript was prepared. I am also grateful for support from the Imperial College Junior Research Fellowship scheme.

\end{document}